\documentclass[reprint,onecolumn,superscriptaddress]{revtex4-2}

\usepackage{url}
\usepackage{graphicx}
\usepackage{bm}
\usepackage[utf8]{inputenc}
\usepackage[T1]{fontenc}
\usepackage{mathptmx}
\usepackage{amssymb}

\begin{document}

\title{A Multipurpose End-Station for Atomic, Molecular and Optical Sciences and Coherent Diffractive Imaging at ELI Beamlines}
\author{Eva Klime\v{s}ov\'a}
\email{eva.klimesova@eli-beams.eu}
\author{Olena Kulyk} 
\thanks{left the project on 30.06.2019}
\author{Ziaul Hoque}
\author{Andreas Hult Roos}
\author{Krishna P. Khakurel}
\author{Mateusz Rebarz}
\author{Matej Jurkovi\v{c}}
\author{Martin Albrecht}
\author{Ond\v{r}ej Finke}
\author{Roberto Lera}
\thanks{left the project on 31.01.2020}
\author{Ond\v{r}ej Hort}
\author{Dong-Du Mai}
\author{Jaroslav Nejdl}
\author{Martin Sokol}
\affiliation{ELI Beamlines, Institute of Physics AS CR, v.v.i., Na Slovance 2, 182 21 Prague 8, Czech Republic}
\author{Rasmus Burlund Fink}
\affiliation{ELI Beamlines, Institute of Physics AS CR, v.v.i., Na Slovance 2, 182 21 Prague 8, Czech Republic}
\affiliation{Department of Physics and Astronomy, Aarhus University, 8000 Aarhus C, Denmark}
\author{Ltaief Ben Ltaief}
\affiliation{Department of Physics and Astronomy, Aarhus University, 8000 Aarhus C, Denmark}
\author{Daniel Westphal}
\affiliation{Laboratory of Molecular Biophysics, Department of Cell and Molecular Biology, Uppsala University, Husargatan 3 (Box 596), SE-75124 Uppsala, Sweden}
\author{Adam Wolf}
\author{Tom\'a\v{s} La\v{s}tovi\v{c}ka}
\affiliation{ELI Beamlines, Institute of Physics AS CR, v.v.i., Na Slovance 2, 182 21 Prague 8, Czech Republic}
\author{Fabio Frassetto}
\author{Luca Poletto}
\affiliation{Institute of Photonics and Nanotechnologies, National Research Council, Via Trasea 7, 35131 Padova, Italy}
\author{Jakob Andreasson}
\author{Maria Krikunova}
\email{maria.krikunova@eli-beams.eu}

\affiliation{ELI Beamlines, Institute of Physics AS CR, v.v.i., Na Slovance 2, 182 21 Prague 8, Czech Republic}

\date{\today}

\begin{abstract}
We report on the status of a users' end-station, MAC: a {\bf M}ultipurpose station for {\bf A}tomic, molecular and optical sciences and {\bf C}oherent diffractive imaging, designed for studies of structure and dynamics of matter in the femtosecond time-domain. MAC is located in the E1 experimental hall on the high harmonic generation (HHG) beamline of the ELI Beamlines facility. The extreme ultraviolet beam from the HHG beamline can be used at the MAC end-station together with a synchronized pump beam (which will cover the NIR/Vis/UV or THz range) for time-resolved experiments on different samples. Sample delivery systems at the MAC end-station include a molecular beam, a source for pure or doped clusters, ultrathin cylindrical or flat liquid jets, and focused beams of substrate-free nanoparticles produced by an electrospray or a gas dynamic virtual nozzle combined with an aerodynamic lens stack. We further present the available detectors: electron/ion time-of-flight and velocity map imaging spectrometers and an X-ray camera, and discuss future upgrades: a magnetic bottle electron spectrometer, production of doped nanodroplets and the planned developments of beam capabilities at the MAC end-station.
\end{abstract}

\maketitle

\section{Introduction}

Ultrafast laser-driven sources based on high-order harmonic generation (HHG) produce light pulses over a broad photon energy range from vacuum ultraviolet to X-rays. Favorable properties of this radiation, namely, femto- to attosecond pulse duration, regular wavefront, high temporal and spatial coherence as well as natural synchronization with the driving laser make HHG sources extremely powerful and versatile tools for applications in physics, chemistry, biology and materials science. Combined research efforts at novel ultrafast light sources based on laser-driven HHG as well as accelerator-driven free electron lasers (FELs) have advanced the field of high intensity laser-matter interactions, atomic, molecular and optical (AMO) sciences and investigations of ultrafast phenomena at short wavelengths \cite{ueda2019,young2018,hochlaf2017,mudrich2014}.

Novel applications of HHG sources in the femtosecond time domain include measurements of electronic structure and dynamics in molecules \cite{nishitani2017,rouzee2014,reid2012,svoboda2019}, mapping the pathways of photochemical reactions \cite{attar2017,smith2018,conta2018,chang2020,warne2020}, coherent diffractive imaging (CDI) both on fixed samples \cite{rothhardt2018,helk2019} and on substrate-free isolated nanoscale samples \cite{rupp2017}, and studies of the interaction of intense extreme ultraviolet (XUV) light with complex systems \cite{murphy2008,bunermann2012,schutte2014,schutte2016}. Time-compensating monochromators allow for the selection of individual harmonics while almost preserving the femtosecond pulse duration \cite{frassetto2011}. This development provides a unique tool for experiments with high temporal and spectral resolution \cite{reid2012,conta2018,warne2020,wernet2011}. A technical challenge for many applications remains in increasing the pulse energy of individual harmonics. Recent progress in the development of driving lasers based on optical parametric chirped pulse amplification (OPCPA) technology enables the up-scaling of the output power of HHG using a loose focusing geometry \cite{hergott2002,takahashi2002,hong2014,heyl2016}.

In addition to HHG sources, which enable novel investigations, advanced sample delivery systems open new possibilities in AMO sciences and CDI. Molecular beams are of interest for chemical physics \cite{smith2018} or for measurements of photoelectron angular distributions in the molecular frame \cite{reid2012}, while cluster/nanodroplet beams are attractive targets for electron spectroscopy \cite{mudrich2014,toennies2004} or for studies of nanoscale matter in extreme conditions \cite{fennel2010,krainov2002,tisch2003}. Liquid targets such as ultrathin liquid sheet jets have enabled investigations of ultrafast radiolysis of water \cite{loh2020} and, due to their high sample velocities, can be used for high-repetition-rate experiments \cite{wiedorn2018}. Moreover, aerosol injectors can deliver a wide range of nanoscale targets ranging from simple sucrose nano-balls \cite{ho2020,rath2014} to viruses and other biological targets for state-of-the-art CDI experiments  \cite{hantke2014,seibert2011,schot2015,sobolev2020}.

ELI Beamlines is a part of the European ELI (Extreme Light Infrastructure) project, developing cutting-edge laser systems for user applications. The L1 Allegra laser system is an in-house development program to deliver 100~mJ laser pulses of $<15$~fs pulse duration at $\sim$830~nm central wavelength and 1~kHz repetition rate \cite{batysta2016}. This laser system can be used to drive a high intensity XUV beamline based on HHG in gas targets \cite{hort2019,nejdl2021}. Here we present the technical design and the current status of the MAC end-station: a \textbf{M}ultipurpose station for \textbf{A}MO sciences and \textbf{C}DI for user experiments with intense HHG pulses. 
We provide an overview of the sample delivery instruments and detection systems, together with results from commissioning experiments. Upcoming upgrades are also discussed.

\section{End-station overview}
\subsection{General layout of the beamline}

\begin{figure*}[t]
	\begin{center}
		\includegraphics[width=15cm]{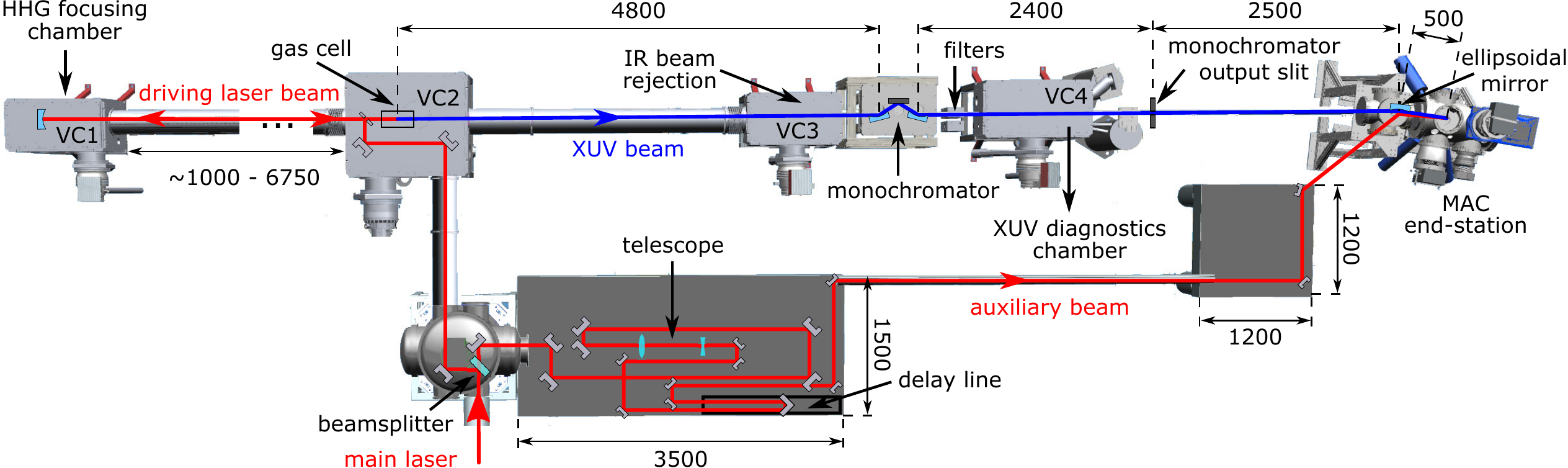}
		\caption{\label{fig_beamline} Layout of the HHG beamline and the MAC end-station. The main laser propagates in a vacuum distribution system from the bottom left. 10\% of the beam is split off and directed out of the vacuum system. The remaining 90\% drives the high harmonic source. The two beams are recombined in the MAC end-station. Dimensions in mm are indicated. Distance between HHG chambers VC1 and VC2 can be modified to accommodate for different focal lengths of the HHG focusing mirror.}
	\end{center}
\end{figure*}
\begin{figure}
	\begin{center}
		{\includegraphics[width=10cm]{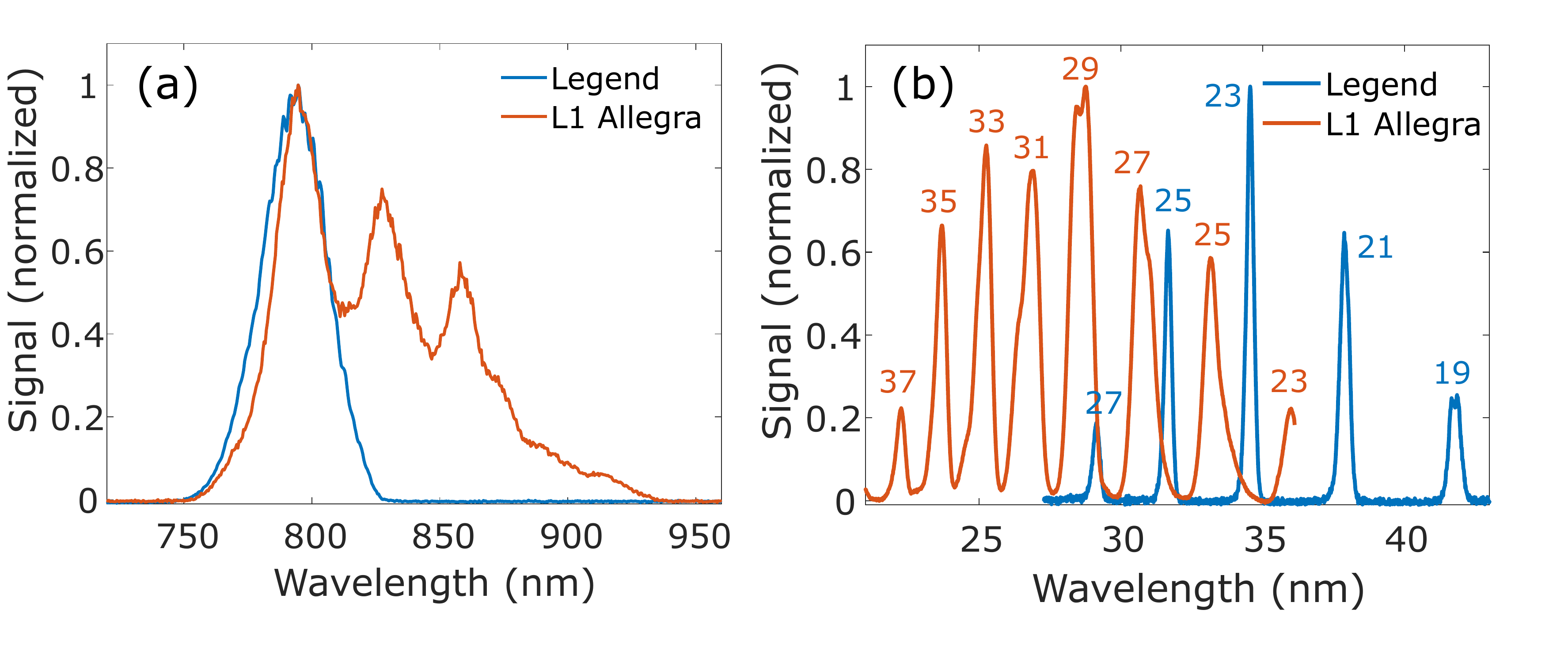}}
		\caption{\label{fig_spectra} (a) Measured spectra of Legend and L1 Allegra lasers. Spectrum of Hidra (not shown) is very similar to the spectrum of Legend. (b) Measured spectra of high-harmonics generated in Kr with Legend (792~nm, 12~mJ, 35~fs, blue line) and with L1 (830~nm, 18.5~mJ, 15~fs, red line). Harmonic orders are indicated. An aluminum filter was used to reject the NIR beam.}
	\end{center}
\end{figure}

The MAC end-station is located at the HHG beamline \cite{hort2019,nejdl2021} of the ELI Beamlines facility (Fig.~\ref{fig_beamline}). The HHG beamline is primarily designed to be driven by the ELI Beamlines L1 laser Allegra \cite{batysta2016}. The Allegra laser system is based on a broadband OPCPA pumped by picosecond Yb:YAG thin-disk lasers. After amplification, the broadband pulses are compressed to $< 15$~fs using a chirped mirror compressor. The system operates at 1~kHz repetition rate with a central wavelength of around 830~nm and an expected energy per pulse of up to 100~mJ. The carrier-envelope phase is not stabilized. Additionally, two commercial support lasers (Legend Elite Duo: $<35$~fs, 12~mJ, 1~kHz, and Hidra-100: $<40$~fs, 100~mJ, 10~Hz, both from Coherent) can be used to drive the beamline. Note that Legend and L1 Allegra have different spectra, resulting in different photon energies of harmonics generated by these lasers (Fig.~\ref{fig_spectra}). 

Before entering the HHG beamline, 10\% of the main laser beam is transmitted through a beamsplitter and directed out of the vacuum beam transport system. This beam serves as an auxiliary beam for pump-probe experiments at the MAC end-station (Section~\ref{sec_aux}). 90\% of the main laser beam is reflected from the beamsplitter and directed to the HHG beamline \cite{hort2019}. In this beamline, the driving laser beam is loosely focused to a gas cell to generate high harmonics in the $5 - 120$~nm wavelength range (photon energies $10 - 250$~eV). Depending on experimental needs, the photon yield in different photon energy ranges can be adjusted by changing the generating gas and f-number of the laser. After generation the XUV beam propagates through the beamline, where the driving NIR beam is rejected by three grazing incidence mirrors (with anti-reflective coating for NIR and high reflectivity for XUV) preserving the original beam axis. A thin metallic filter (typically aluminum, zirconium or indium) can be used to block the residual NIR beam and transmit a specific part of the high harmonic spectrum. Transmission of the beam-rejection system and filters has to be taken into account when designing experiments in a specific photon energy range.

For experiments at the MAC end-station that require a monochromatized beam, a time-preserving grating monochromator is implemented \cite{frassetto2011,hort2019,frassetto2017,poletto2018}. The monochromator consists of two toroidal mirrors and a grating in an off-plane diffraction geometry. With this geometry the wavefront tilt is less pronounced than in a classical diffraction mount \cite{frassetto2017,poletto2018}, see Table~\ref{tab_gratings}. Slots for four gratings are mounted on a motorized stage in the monochromator vacuum chamber allowing a quick exchange of a grating. Altogether, six gratings with parameters listed in Table~\ref{tab_gratings} and a flat golden mirror (if a broadband beam is needed) are available. A slit at the monochromator output selects the specific part of the spectrum to be used in the experiment. One of five slits (with widths of 50, 80, 100, 200 and 400~$\mu$m) can be selected. If the monochromator is not needed its components can be moved out of the XUV beampath.

\begin{table}
	\caption{\label{tab_gratings}Gratings available for the monochromator. Energy resolution $\Delta E$ and pulse temporal front tilt (half-width) $\Delta\tau$ after diffraction are calculated for a beam with full divergence of 0.5~mrad at a photon energy in the center of the optimal range for each grating. A slit width of 100~$\mu$m was considered for the $\Delta E$ calculation \cite{frassetto2011}.}
	\begin{center}
	\begin{tabular}{c|cccc}
		\hline\noalign{\smallskip}
		
		Grating&Lines/mm&\begin{tabular}{c} Optimal spectral \\ range (eV)\\\end{tabular}&$\Delta E$~(eV)&$\Delta\tau$~(fs)\\
		
		\noalign{\smallskip}\hline\noalign{\smallskip}
		
		1&86&$10-28$&$0.14$&$22$\\
		2&150&$13-28$&$0.10$&$35$\\
		3&158&$25-54$&$0.34$&$20$\\
		4&300&$22-40$&$0.11$&$48$\\
		5&600&$51-98$&$0.31$&$40$\\
		6&985&$86-120$&$0.36$&$47$\\
		\noalign{\smallskip}\hline
	\end{tabular}
\end{center}
\end{table}

The default polarization of the XUV beam in the MAC end-station is vertical and monochromator transmission is optimized for this polarization. Nevertheless, it is possible to rotate the polarization of the driving laser field using a thin half-waveplate or a periscope and thus rotate the polarization of the XUV beam. In the future, circular and elliptical polarization of the XUV beam will also be available, employing a two-color driving scheme for the HHG \cite{kfir2015}.

A major planned upgrade of the HHG beamline is to provide two independent synchronized XUV beams for pump-probe experiments. 

\subsection{MAC vacuum chamber}
The design of the MAC vacuum chamber is based on the design of the CAMP end-station at the FLASH FEL in Hamburg \cite{struder2010,erk2018} and the LAMP instrument at the Linac Coherent Light Source FEL \cite{osipov2018}. This provides mutual compatibility of the chambers, eg. instruments used at CAMP can easily be integrated into the MAC end-station. The main body of the MAC chamber is a DN400CF cylinder made of stainless steel 1.4429~ESU (316LN~ESR) to ensure low magnetic permeability. The interaction region is in the center of the first part of the chamber, as indicated in Figs.~\ref{fig_focusing}(a) and \ref{fig_pump_beam}(a). There are four DN250CF ports around the interaction region and a number of other ports on the chamber. For installation of optical components inside the MAC chamber, three optical breadboards are available: one at the top and two at the bottom part of the DN400CF cylinder. The flexible, multi-purpose configuration of the MAC station allows for the integration of user's instruments on request.

The MAC chamber is located in an ISO-7 class cleanroom (experimental hall E1) with an ambient temperature of $20^{\circ}{\rm C} - 22^{\circ}{\rm C}$, a temperature stability of $\pm 0.5^{\circ}$C over a 24 hour period and a humidity of $50\% \pm 5\%$. The MAC chamber is pumped by a turbomolecular pump with a pumping speed of 2100~l\,s$^{-1}$, reaching a vacuum level of 10$^{-8}$~mbar. Additional turbomolecular pumps or cold traps can be added if required.

\subsection{XUV beam focusing at MAC end-station}

\begin{figure}
	\begin{center}
		\includegraphics[width=8.5cm]{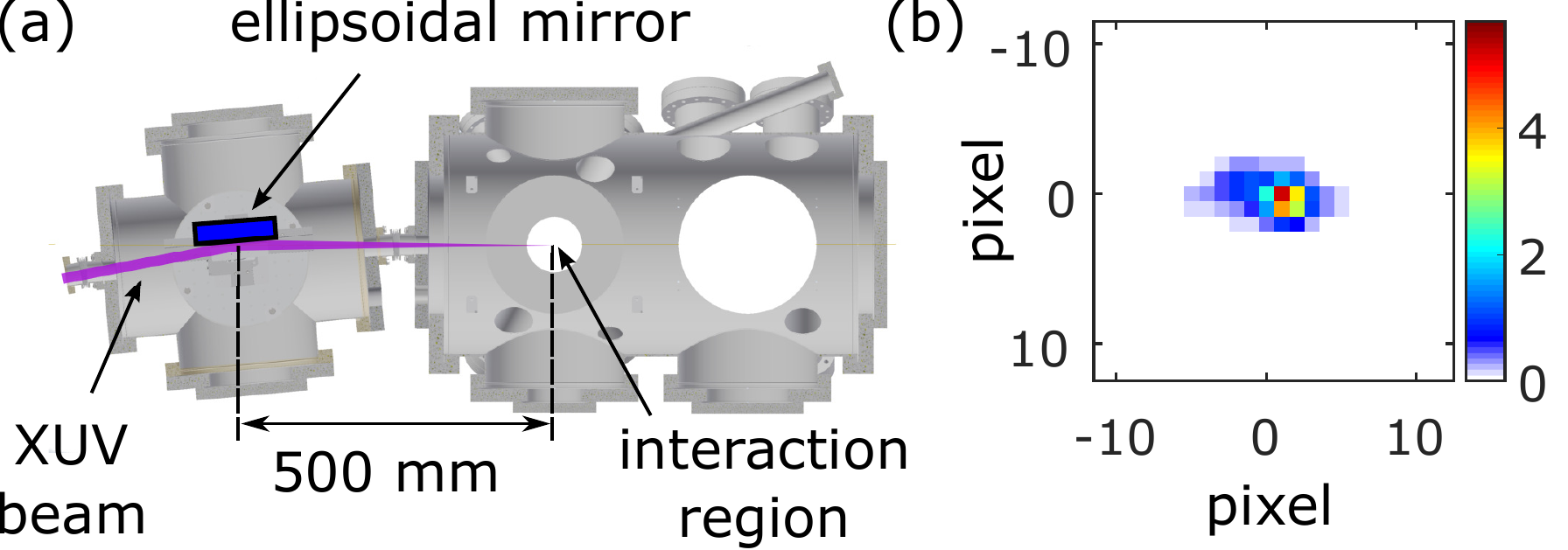}
		\caption{\label{fig_focusing} HHG beam focusing to the MAC end-station. (a) An ellipsoidal mirror images the output slit of the monochromator to the interaction region in the MAC chamber. (b) Measured focal spot of the 21st harmonic focused with the ellipsoidal mirror. The focal spot was measured with an in-vacuum CCD camera (PI-MTE) with a pixel size of 13.5~$\mu$m.}
	\end{center}
\end{figure}

The XUV beam can be focused to the MAC end-station in two main ways. In the first focusing geometry, the output slit of the monochromator is imaged with an ellipsoidal mirror to the MAC chamber with a 1:5 imaging ratio (Fig.~\ref{fig_focusing}(a)). The focusing mirror, located in a separate chamber in front of the MAC chamber, is a gold-coated ellipsoidal mirror with an entrance arm of 2500~mm, an exit arm of 500~mm and a grazing angle of an incidence of 5~degrees. This geometry has been implemented and the measured focal spot of 21st harmonic (photon energy 33~eV) is shown in Fig.~\ref{fig_focusing}(b). In this measurement, a grating with 150~lines/mm was used in the first diffraction order and the slit size was 200~$\mu$m. The vertical full width at half maximum (FWHM) of the measured focal spot is 40~$\mu$m in accordance with the 1:5 imaging ratio. The FWHM of the focal spot in the horizontal direction ($\sim$60~$\mu$m) corresponds to the width of the harmonic beam, not affected by the slit.

To achieve tight focusing, a second geometry employing an off-axis parabola (OAP) can be implemented. The OAP has a focal length of 268.94~mm and an angle between the incoming and the focused beam of $\sim 21.5^{\circ}$.  In this configuration, the expected focal spot size in the MAC end-station is below $3\;\mu$m (FWHM). With the corresponding estimated intensity on the target of above $10^{13}$~W\,cm$^{-2}$ we expect to meet the experimental requirements for single-shot single-harmonic CDI and for studies of non-linear effects in the XUV regime \cite{nayak2018,senfftleben_2020}.

\subsection{Auxiliary beam for MAC end-station}
\label{sec_aux}
\begin{figure}[t]
	\begin{center}
		\includegraphics[width=15cm]{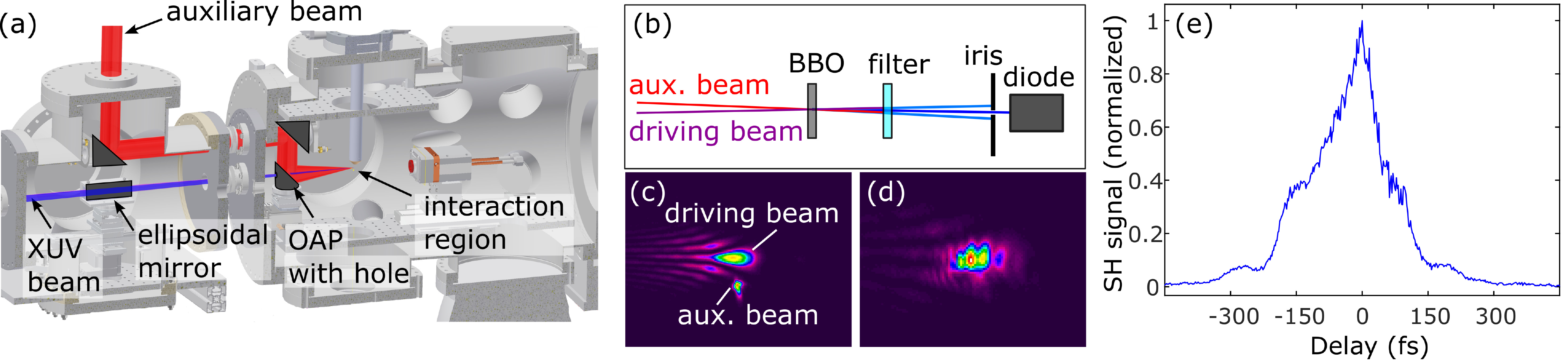}
		\caption{\label{fig_pump_beam} (a) Model of the XUV (blue) and auxiliary (red) beams focused colinearly in the MAC end-station.(b)  Schematic of the experimental setup for non-colinear second-harmonic (SH) cross-correlation. Spatial profiles of the driving and auxiliary beam: (c) not overlapped, (d) overlapped in space and time in the interaction region. Interference fringes are visible. Camera pixel size 4.4~$\mu$m. (e) SH cross-correlation trace of the auxiliary and driving beam.}
	\end{center}
\end{figure}

For pump-probe experiments, an auxiliary beam synchronized with the XUV beam, is provided at the MAC end-station. The auxiliary beam is split from the main laser beam before the HHG beamline (Fig.~\ref{fig_beamline}) and propagates in air to the MAC end-station. The auxiliary beam passes through a delay line with a total travel of 1~m and a bi-directional repeatability of $<0.5~\mu$m (Newport, M-IMS1000LM-S), providing a total delay of up to 6~ns with a delay step of $<3$~fs. After the delay line, the auxiliary beam is directed to the MAC chamber and focused on the interaction point. Different focusing geometries (colinear, non-colinear, using a lens or an OAP) are possible (examples are shown in Fig.~\ref{fig_pump_beam}(a) and \ref{fig_pump_probe}(a)). 

The spatial and temporal overlap of the driving and the auxiliary beam is found by propagating only the driving NIR beam on the same path as the XUV beam. The auxiliary beam and the driving beam are imaged on a camera in the interaction region to ensure their spatial overlap. For a rough temporal overlap (with an uncertainty of few picoseconds) a fast photodiode is used. Second harmonic (SH) generation in a non-linear crystal (BBO) is used to find the zero delay between the two beams with $<3$~fs uncertainty (Fig.~\ref{fig_pump_beam}(b-e)). The maximum of the SH cross-correlation signal (Fig.~\ref{fig_pump_beam}(e)) determines the zero delay between the two pulses. 

The FWHM of the SH cross-correlation signal is around 135~fs, which is much larger than the 35~fs initial pulse duration. This is because the auxiliary beam undergoes a large amount of positive dispersion on the way to the MAC chamber. An implementation of chirped mirrors is planned to post-compress the auxiliary beam to a pulse duration of $< 35$~fs.

Upgrades of the auxiliary beam are planned to cover a wide spectral range. A high-energy optical parametric amplifier (OPA, Light Conversion, HE-TOPAS + ViS/UV extension) will be installed to produce radiation in the wavelength range of $240-2600$~nm. Additionally, THz radiation can be generated on the auxiliary beampath. Generation of single-cycle broadband $\sim 1$~THz pulses will be realized in the tilted wave-front geometry in a LiNbO$_3$ crystal \cite{yeh2007,hebling2008}. To generate a narrow-band high-intensity tunable THz field, a setup based on difference frequency generation of two pulses in a DAST crystal \cite{liu2017} is planned.

\subsection{Pump-probe setup for XUV and auxiliary beam}

\begin{figure}[t]
	\begin{center}
		\includegraphics[width=15cm]{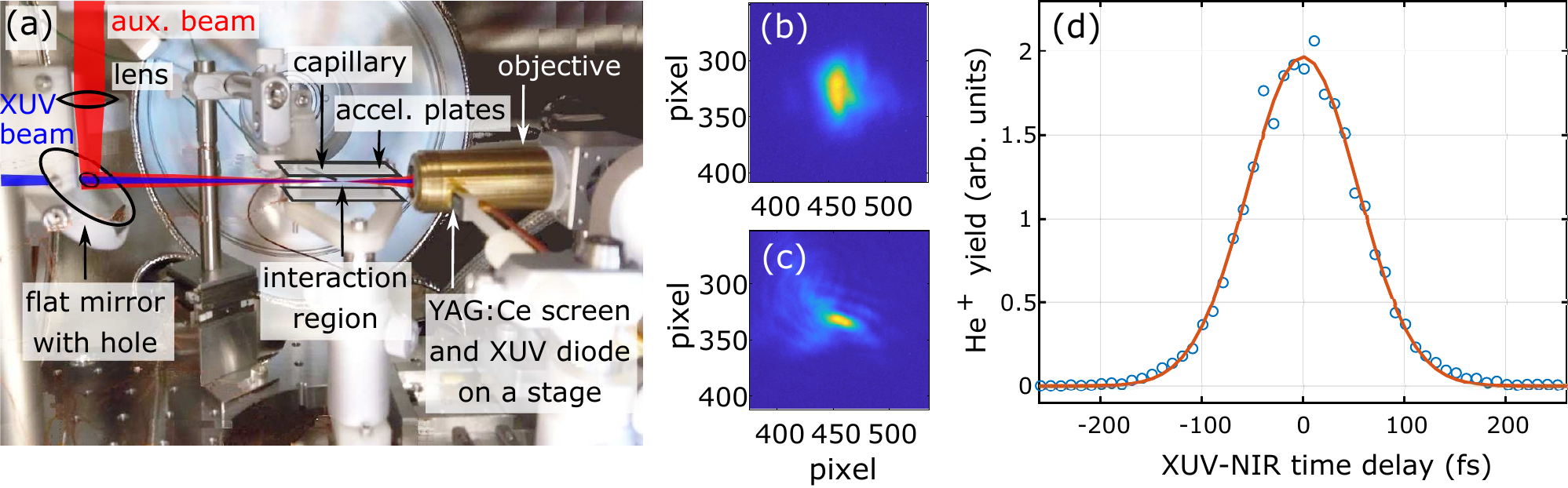}
		\caption{\label{fig_pump_probe} (a) Photograph of the setup to determine spatial and temporal overlap of the XUV and auxiliary beams. The beams are focused colinearly into the interaction region. Ions are accelerated by the electrical field of 100~V\,mm$^{-1}$ between two accelerator (accel.) plates and detected with a microchannel plate detector (not shown). (b) 5$\times$ magnified image of the focal spot of the 15th harmonic on a YAG:Ce screen. (c) 5$\times$ magnified image of the auxiliary beam focal spot at the same position as the XUV beam focus. Camera pixel size 3.75~$\mu$m. (d) Transient He$^+$ yield measured as a function of the delay between XUV and NIR pulses (15th harmonic, photon energy 23.5~eV, 10$^7$~photons/shot, NIR peak intensity $10^{13}$~W\,cm$^{-2}$) with time-delay steps of 10~fs. Each data point is an average over 3000 single shots. Red solid line is a fit with the Gaussian function.}
	\end{center}
\end{figure}

Many experiments at the MAC end-station require synchronized XUV and auxiliary beams. A pump-probe setup for XUV and NIR pulses is shown in Fig.~\ref{fig_pump_probe}(a). To ensure a colinear focusing geometry, the XUV beam is focused by the ellipsoidal mirror and passes through a flat mirror with a hole. The auxiliary beam is coming from the top. It is focused with a lens and then recombined with the XUV beam using the flat mirror with the hole. A YAG:Ce fluorescence screen, mounted on a motorized stage, can be placed into the interaction region to image the XUV focal spot. The interaction plane is imaged with an infinity-corrected microscope objective (magnifications of 2$\times$ and 5$\times$ are available) on a camera (Manta G-125B, pixel size 3.75~$\mu$m) placed outside the MAC chamber (Fig.~\ref{fig_pump_probe}(b,c)). Spatial overlap is ensured by centering the two focal spots on the same position on the camera. The mirror with the hole is mounted on a motorized mount allowing fine-tuning of the spatial overlap of the beams. Besides the YAG:Ce screen, a calibrated XUV photodiode can be moved into the XUV beam to determine the photon flux in the interaction region.

The temporal overlap of the XUV and auxiliary beams is found by detecting ions created by the ionization of an atomic gas in combined XUV and NIR fields. A metal capillary with inner diameter of 0.88~mm, placed few millimeters from the interaction region, is used to deliver a gas target. Ions created in the interaction are detected by an ion time-of-flight spectrometer (section~\ref{sec_tof}). Alternatively, velocity map imaging optics (section~\ref{sec_vmi}) can be used. Fig.~\ref{fig_pump_probe}(d) shows the transient He$^+$ yield measured as a function of delay between the 15th harmonic of the Legend laser (photon energy 23.5~eV, 10$^7$~photons/shot) and an auxiliary beam with a peak intensity of $10^{13}$~W\,cm$^{-2}$. There is no He$^+$ signal in the presence of only one beam, because the 15th harmonic photon energy of 23.5~eV is below the ionization energy of He (24.6~eV) and the NIR beam intensity is not sufficient to tunnel-ionize He. He$^+$ ions are created by non-sequential two-color ionization \cite{bottcher2007} in combined harmonic and NIR pulses. The FWHM of the cross-correlation trace is 130~fs.

\section{Sample delivery systems}

The MAC end-station is primarily designed for experiments on low-density targets, such as atoms, molecules, clusters or aerosols. In this section, sample delivery systems at the MAC end-station are reviewed. Besides low-density targets, it is also possible to install solid samples into the MAC chamber. Piezo-driven manipulators with linear (xyz) degrees of freedom as well as sample rotation are available for fine alignment of fixed targets. 

\label{sec_samples}
\subsection{Molecular and cluster beams based on pulsed valves}
\label{sec_mol_beams}

Beams of cold molecules or clusters are of interest for chemical physics, fundamental investigations in AMO sciences \cite{reid2012} or for spectroscopic studies of pure or doped clusters and nanodroplets \cite{mudrich2014,toennies2004}. These beams are produced by gas expansion into a vacuum. Typically a pulsed valve is used for gas expansion into a vacuum in order to achieve high flux in the droplet beam \cite{mudrich2014} while limiting the requirements on the pumping speed.

\begin{figure}
	\begin{center}
		\includegraphics[width=15cm]{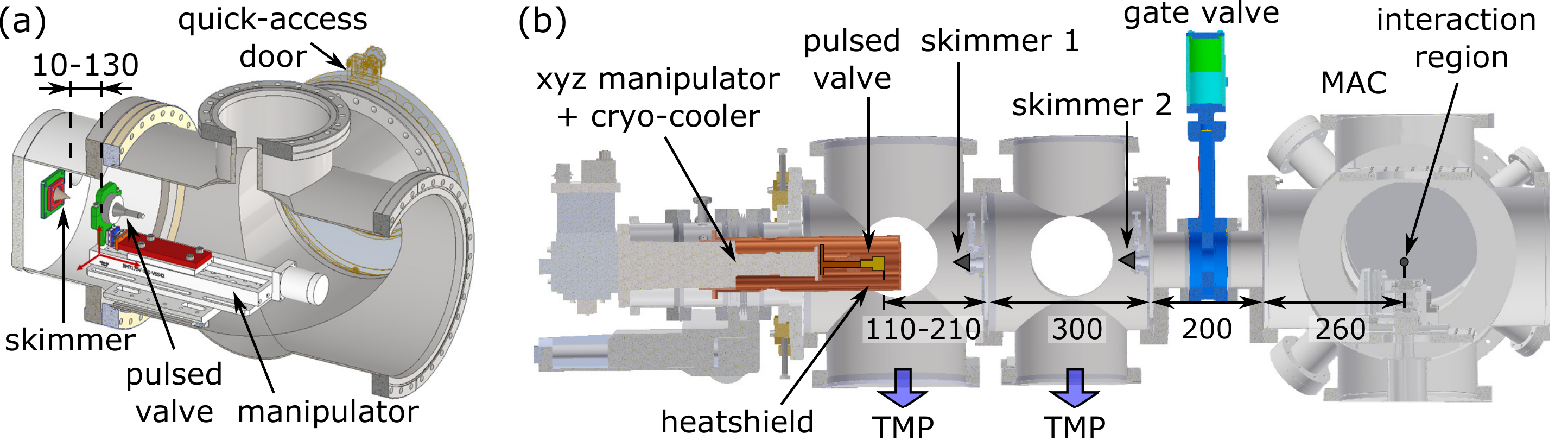}
		\caption{\label{fig_cluster_source} (a) Model of the molecular beam setup. Gas is expanded to vacuum through a pulsed valve. A skimmer selects the central part of the molecular beam. (b) Model of the cluster source based on a cryo-cooled Even-Lavie valve. Dimensions in mm are indicated. TMP -- turbomolecular pump.}
	\end{center}
\end{figure}

Two instruments based on pulsed valves are available at the MAC end-station: a molecular beam source (Fig.~\ref{fig_cluster_source}(a)) and a cluster source (Fig.~\ref{fig_cluster_source}(b)).
In the molecular beam setup (Fig.~\ref{fig_cluster_source}(a)) the gas expands through a valve (Amsterdam Piezo Valve, model ACPV2-100) equipped with a conical 100~$\mu$m diameter nozzle with an opening angle of 40$^{\circ}$. The valve operates at room temperature with backing pressures of up to 30~bar and a repetition rate of up to 5~kHz. It produces gas pulses with a duration of 20~$\mu\mathrm{s} - 1$~ms (or continuous) and with $>10^{16}$ particles per pulse. The pulsed valve is mounted on a stage with three degrees of freedom (xyz), while a skimmer (Beam Dynamics, available diameters between $0.2-3$~mm) is placed on a fixed mount. The distance between the skimmer and the interaction region is 195~mm and the nozzle-skimmer distance can be varied from 10~mm to 130~mm. The vacuum chamber for the molecular beam source is pumped by a 2100~l\,s$^{-1}$ turbomolecular pump. Additional pumping can be added if required.

The second instrument using a pulsed valve is dedicated for the production of clusters or helium nanodroplets (Fig.~\ref{fig_cluster_source}(b)).  The gas expands through a cryo-cooled Even-Lavie valve \cite{even2015} that can operate at a maximum repetition rate of 500~Hz, with a typical opening time of $15-30~\mu$s, a backing pressure up to 100~bar, and operating temperatures down to 4~K. Low temperatures are achieved using a two-stage helium cryo-cooler (Sumitomo, RDK-408D2). The nozzle is trumpet shaped with a $70 - 80\;\mu$m hole and a 40$^{\circ}$ cone angle (Even-Lavie $\#$ 2-70-T-Ry Sapphire (Ruby)). After the gas expands, it propagates through two skimmers (Beam Dynamics) to ensure efficient differential pumping of the vacuum chambers. Both skimmers are mounted on 2-axes motorized stages to allow their alignment under vacuum. The distances between valve, skimmers and the interaction region are indicated in Fig.~\ref{fig_cluster_source}(b). The Even-Lavie valve is mounted on an xyz manipulator, thus the distance between the valve and the first skimmer can be adjusted in the range of $\sim110-210$~mm. The whole setup is separated from the main MAC chamber by a gate valve with a window. The cluster source setup is pumped by two turbomolecular pumps with pumping speeds of 2200 and 1600~l\,s$^{-1}$, respectively.

An upgrade of the cluster apparatus will introduce the capability to dope clusters with atoms or molecules. For this purpose a doping stage can be added to the second vacuum chamber (between the two skimmers). A leak valve is installed in this chamber to provide a source of atoms that can be picked up by the helium droplets. Additionally, an oven will be used to evaporate atoms or molecules of different species that can attach to helium droplets. Altogether, the cluster and the molecular beam sources provide important targets for investigations in AMO sciences.

\subsection{Microfluidic liquid sheets and jets}

Free-flowing liquid jets, both cylindrical \cite{deponte2008,nelson2016} or sheet jets \cite{koralek2018}, are available for use at the MAC end-station. 

Ultrathin cylindrical liquid jets can be produced by a gas dynamic virtual nozzle (GDVN) \cite{deponte2008,muhlig2019}. A GDVN system consists of two concentric capillaries: an inner capillary for the sample liquid and an outer capillary for a carrier gas (typically helium). The coaxial gas dynamic forces compress the liquid jet, leading to a jet with a diameter much smaller than the solid nozzle diameter. GDVNs are generally manufactured either from glass capillaries \cite{deponte2008} or by 3-dimensional printing with sub-micrometer resolution \cite{nelson2016}. The GDVNs available at ELI Beamlines have been manually fabricated at Uppsala University. They typically operate with flow rates in the range of $0.5-5\;\mu$l\,min$^{-1}$ and carrier gas pressures around $15-30$~bar, producing liquid jets with $0.3-1\;\mu$m diameter.

\begin{figure}
	\begin{center}
		\includegraphics[width=7cm]{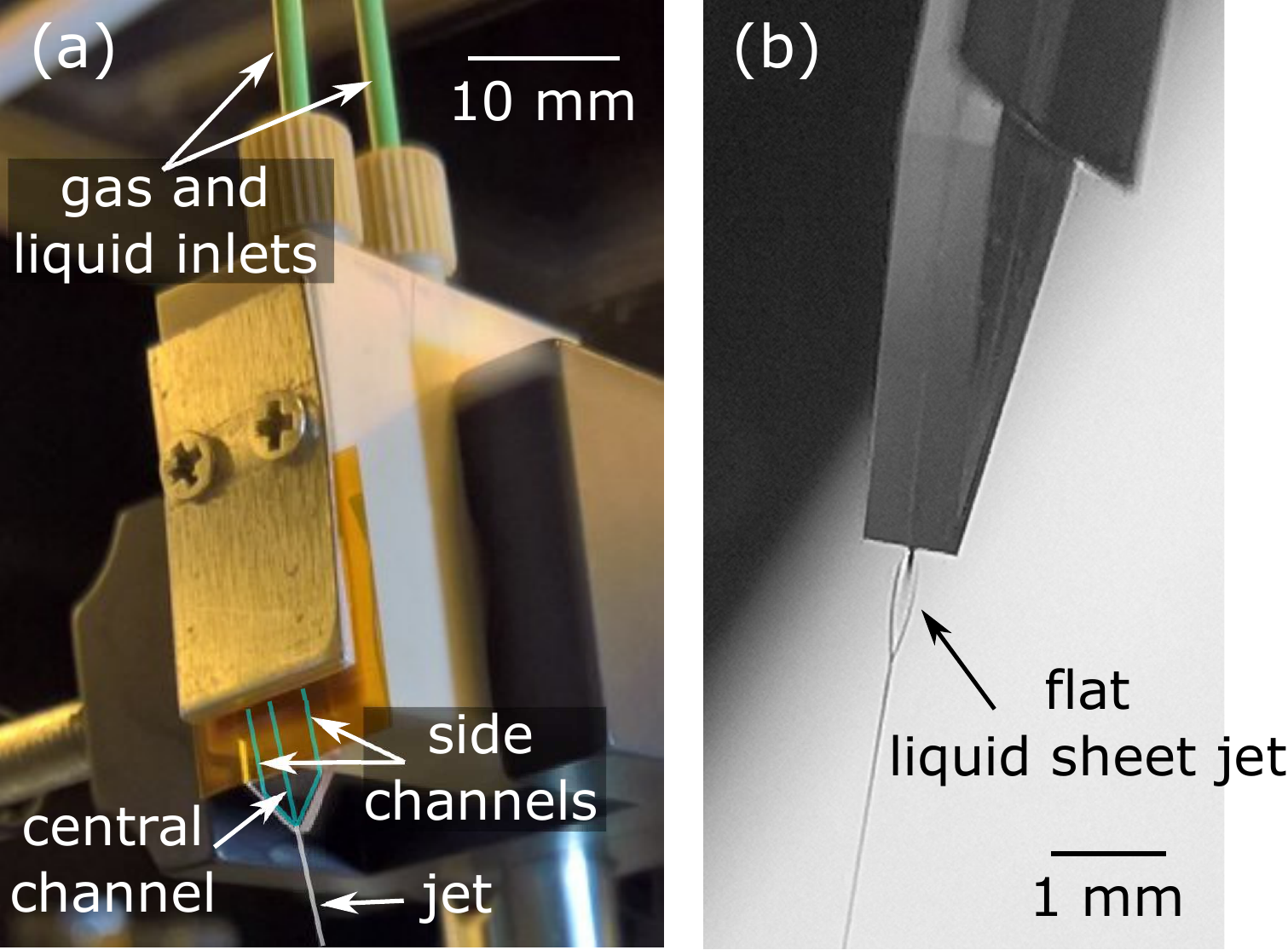}
		\caption{\label{fig_sheets} (a) Vacuum-compatible microfluidic sheet jet assembly during operation. (b) Flat liquid sheet jet produced from a nozzle made of two 0.4~mm-thick borosilicate glass wafers etched together. The width of the water sheet at its widest part is around 200~$\mu$m.}
	\end{center}
\end{figure}

Ultrathin sheet jets at ELI Beamlines have been developed as part of a collaboration with LCLS SLAC \cite{koralek2018}. The liquid sheet jets are generated using a microfluidic glass chip (Fig.~\ref{fig_sheets}) that contains three microfluidic channels: two side channels for gas and a central channel for liquid. The liquid is compressed by two colliding gas jets to form a thin sheet. Using gas dynamic forces instead of colliding liquid jets \cite{ekimova2015,george2019} leads to a lower flow rate and reduced sheet thickness. Under typical operating conditions (liquid flow rate $0.1-0.5$~ml\,min$^{-1}$ and a gas flow rate up to $50-300$~ml\,min$^{-1}$), the thickness of the liquid sheet can range from $> 1~\mu$m  down to $\sim 20$~nm. The flat sheet length is typically on the order of 100~$\mu$m with a width between 10 and 100~$\mu$m.

Although the microfluidic chip has been primarily designed for ultrathin liquid sheet jet generation, it has two alternative modes of operation. When the liquid is running through the central channel and side channels are not used, a cylindrical jet with diameter of around $20-30\;\mu$m is formed. Alternatively, when liquid is injected through the two side channels and the central channel is unused, a colliding sheet jet with a thickness in the range of $3-10\;\mu$m is produced \cite{schulz2018}.

The liquid sheet jet system at ELI Beamlines has been commissioned in air. However, based on experiments performed elsewhere, we expect the liquid sheet jet to operate stably at a vacuum level of 10$^{-3}$~mbar \cite{koralek2018}. The vacuum level can be further improved by employing a differentially-pumped shroud around the liquid jet source or using a cold trap.

\subsection{Aerosol nanoparticle injectors}
\label{sec_injector}

\begin{figure}
	\begin{center}
		\includegraphics[width=8cm]{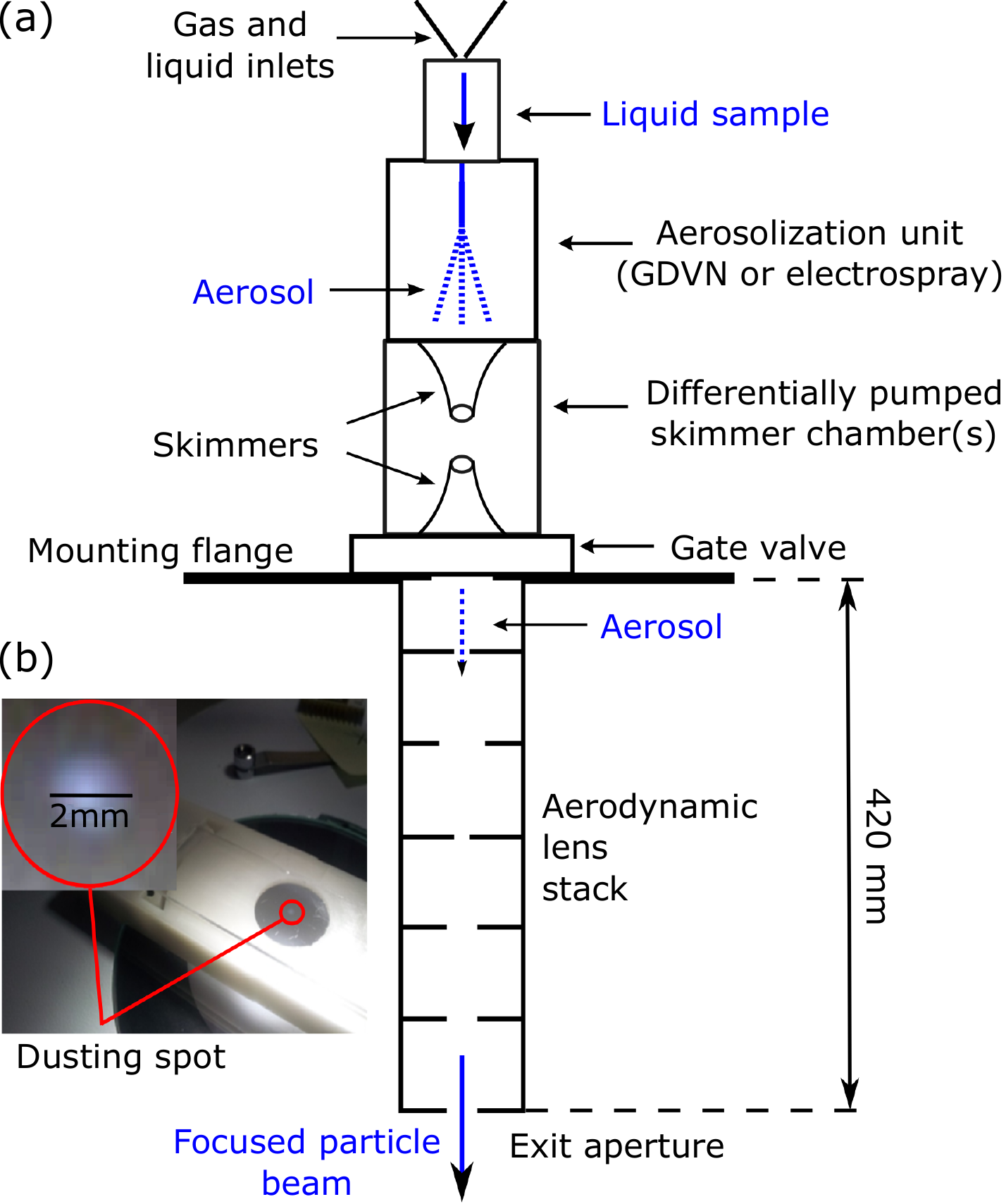}
		\caption{\label{fig_injector} (a) Schematic diagram of the aerosol injector comprising of an aerosolization unit (virtual gas dynamic nozzle (GDVN) or an electrospray (ES) system), a differentially pumped skimmer chamber (or two skimmer chambers for ES) and aerodynamic lens stack  for focusing particles. (b) Dusting test with CsI nanoparticles at ELI Beamlines. Top left -- microscope image of the dusting spot on a glass covered with grease.}
	\end{center}
\end{figure}

The aerosol nanoparticle injector, the so-called “Uppsala injector”,  was originally developed at Uppsala University to deliver a focused beam of substrate-free nanoparticles, biomolecules, viruses or cells to vacuum for CDI with FELs \cite{ho2020,rath2014,hantke2014,seibert2011,schot2015,sobolev2020,andreasson2014,hantke2018,bielecki2019}. Two types of sample aerosolization for the injection can be used: a GDVN \cite{deponte2008,hantke2018} or an electrospray (ES) unit \cite{bielecki2019,chen1995,ganan-calvo2009}. Both are available at ELI Beamlines.

Using the injector with a GDVN, an aerosol is created from the liquid sample as the jet from the GDVN breaks up into droplets. The aerosol propagates through a skimmer chamber for excess gas removal (Fig.~\ref{fig_injector}).  After that it enters an aerodynamic lens stack (ALS) \cite{wang2005,wang2006,liu2007}, consisting of a set of apertures that guide and focus the nanoparticle beam into the MAC chamber. Water from the sample solution is evaporated on the way to the vacuum chamber and dry container-free nanoparticles are delivered for the interaction with the laser. Nanoparticles with sizes in the range of $\sim70-2000$~nm can be injected by the injector with the
GDVN and the focused particle beam waist can have a diameter down to $\sim10~\mu$m \cite{hantke2018}. The volume density of the injected nanoparticles is typically of the order of $10^8$~cm$^{-3}$.

The injection at ELI Beamlines has been verified by a dusting test of the injected nanoparticles (Fig.~\ref{fig_injector}(b)). A 0.1\% water solution of CsI salt was injected onto a microscope glass covered with vacuum grease, located 25~mm from the ALS tip. After the injection, a white dusting spot on the glass was visible by eye, confirming injection of CsI to the chamber (Fig.~\ref{fig_injector}(b)). For a more accurate determination of particle sizes and velocities a Rayleigh scattering setup is under development \cite{hantke2018}.

\begin{figure}
	\begin{center}
		\includegraphics[width=8cm]{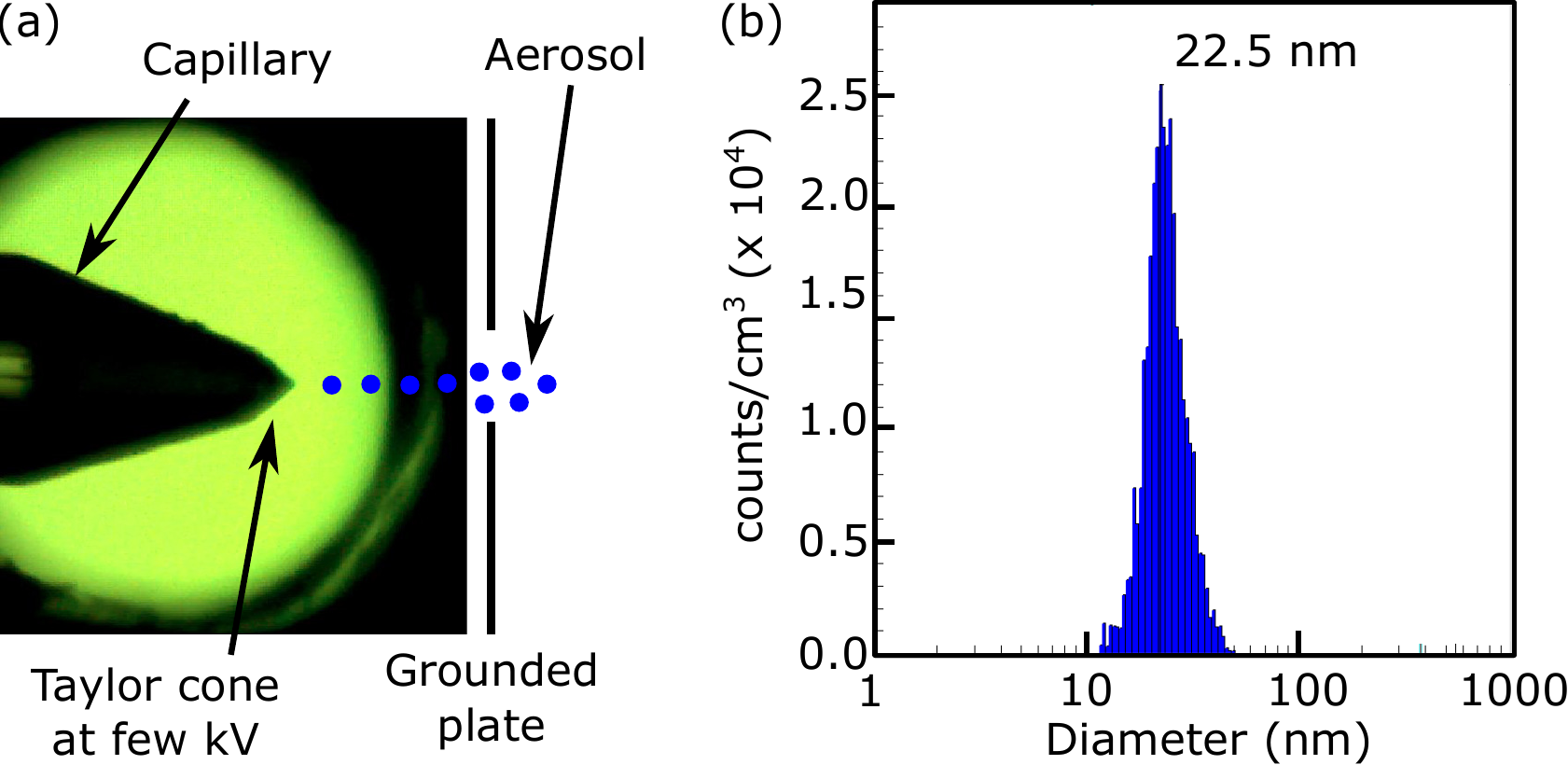}
		\caption{\label{fig_electrospray} Electrospray injection at ELI Beamlines. (a) Taylor cone created by electrospraying a 10\% sucrose solution in ammonium acetate. (b) Histogram of sucrose particle sizes measured with the differential mobility analyzer. Mean diameter is 22.5~nm.}
	\end{center}
\end{figure}

The second aerosolization system available for the MAC end-station is an electrospray unit \cite{chen1995,ganan-calvo2009}. For ES aerosolization, the sample is prepared in a conducting solvent (typically ammonium acetate) and a voltage of about $2 - 3$~kV is applied to it. The sample liquid flows through a capillary placed close to a grounded plate and the liquid jet from the capillary is squeezed by electrostatic forces into a Taylor cone (Fig.~\ref{fig_electrospray}(a)). The tip of the Taylor cone breaks into charged droplets, which are neutralized by an X-ray source, and then enter an injector setup consisting of two skimmer chambers and an ALS, similar to that of the injector with the GDVN (Fig.~\ref{fig_injector}(a)). The injector with ES unit is suitable for injection of (bio)particles with diameters down to $\sim$10~nm \cite{bielecki2019}.

The operation of the ES unit was commissioned with a differential mobility analyzer (DMA, TSI Incorporated). For the DMA measurement, the output of the ES unit is connected with a conductive resin tube to the DMA instrument, where nanoparticles drift in an applied electric field and their diameters are determined from their electrical mobility. Results of the size measurement of sucrose nanoparticles (1\% sucrose concentration in 25~mM ammonium acetate) are shown in Fig.~\ref{fig_electrospray}(b). The mean diameter was found to be 22.5~nm, which is within the expected size range for this concentration.

The injector, either with GDVN or ES, can be installed on the top DN250CF port of the MAC chamber on an xyz manipulator. When the injector is in operation, the vacuum level in MAC chamber rises typically to $10^{-6}-10^{-5}$~mbar, still allowing ion or electron spectroscopy \cite{klimesova2019} (section~\ref{sec_tof}, Fig.~\ref{fig_itof}(b)).

\section{Charged particle spectrometers and photon detectors}

\label{sec_detectors}

\subsection{Electron and ion time-of-flight spectrometer}
\label{sec_tof}

Electrons or ions created during interactions in the MAC end-station can be detected by a linear electron or ion time-of-flight (ToF) spectrometer. Kinetic energies of electrons and the mass-to-charge ratio of ions are determined from their flight times. To increase the acceptance solid angle of electrons, electrostatic lenses can be used. The detector is a 40~mm diameter microchannel plate (MCP) coupled to an anode. Data from the anode can be acquired at 1~kHz repetition rate on a single-shot basis using a digitizer (SP Devices, 10~GSa/second, 14~bit resolution). 

\begin{figure}[t]
	\begin{center}
		\includegraphics[width=15cm]{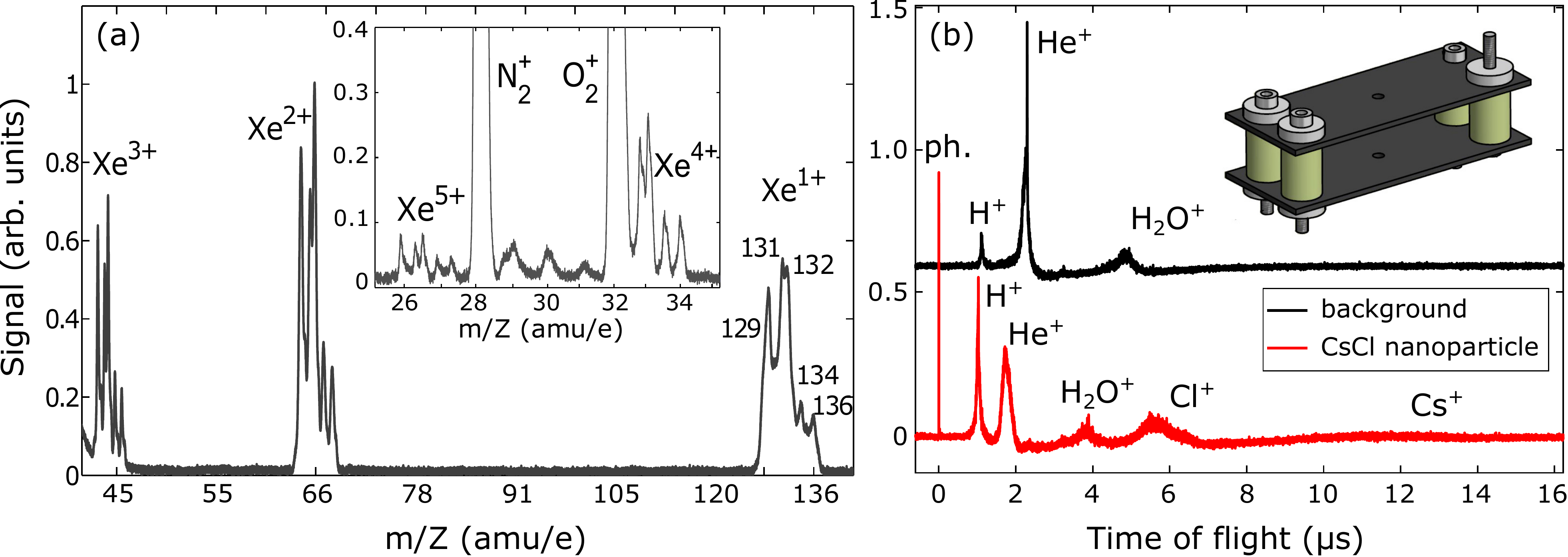}
		\caption{\label{fig_itof} (a) Measured ion time-of-flight trace from xenon gas (chamber backfilled at $7.4\times10^{-7}$~mbar) irradiated by an intense laser beam (800~nm, 120~fs, $2\times10^{15}$~W\,cm$^{-2}$). Five Xe isotopes are resolved. (b) Measured single-shot ion time-of-flight traces from injected CsCl nanoparticles (diameter $\sim$100~nm) irradiated by an intense laser beam (800~nm, 120~fs, $2\times10^{15}$~W\,cm$^{-2}$). Traces are presented with a vertical offset for clarity. Top trace: background and carrier gas. Bottom trace: CsCl nanoparticle, (ph. stands for photon peak). Inset: 3D model of electrostatic plates for ion acceleration.}
	\end{center}
\end{figure}

For detection of ions, a set of plates with an applied voltage for ion acceleration, mounted on a piezo-driven xyz stage, can be used (see inset in Fig.~\ref{fig_itof}(b) as an example). Measured ion ToF traces from a xenon gas (chamber backfilled at $7.4\times10^{-7}$~mbar) irradiated with an intense 800~nm laser beam (pulse duration 120~fs, peak intensity $2\times10^{15}$~W\,cm$^{-2}$) are displayed in Fig.~\ref{fig_itof}(a).  Xe ions with charges up to 5+ can be observed, which is expected for the laser intensity used. Five Xe isotopes with mass numbers $A = 129$, 131, 132, 134 and 136, respectively, can be well resolved in all charge states, confirming sufficient mass resolution of the ion ToF spectrometer.

The ion ToF spectrometer can be used together with the nanoparticle injector (section~\ref{sec_injector}).
In the first commissioning experiments, the gas background from the nanoparticle injector has been thoroughly characterized by the determination of energies of ions created in the interaction with a strong NIR laser field. It has been shown that a plasma channel can be created in the interaction volume and the measured ion energies can be used to estimate the gas density \cite{klimesova2019}.

Using this setup, we have also performed ion spectroscopy on injected CsCl nanoparticles (diameter $\sim$100~nm) irradiated by a strong laser beam (wavelength 800~nm, pulse duration 120~fs, peak intensity $2\times10^{15}$~W\,cm$^{-2}$), as shown in Fig.~\ref{fig_itof}(b). Ions were detected for single laser shots. When a nanoparticle is hit, Cl$^+$ and Cs$^+$ ions from the nanoparticle and a photon peak appear in the ion ToF spectrum. Cl$^+$ and Cs$^+$ peaks are rather broad because they contain ions with different kinetic energies, coming from the explosion of the laser-irradiated nanoparticle.
This measurement demonstrates the possibility to perform ion ToF spectroscopy on different types of nanoparticles injected into the vacuum chamber using the aerosol injector.

\subsection{Velocity map imaging spectrometer}
\label{sec_vmi}

\begin{figure}
	\begin{center}
		\includegraphics[width=15cm]{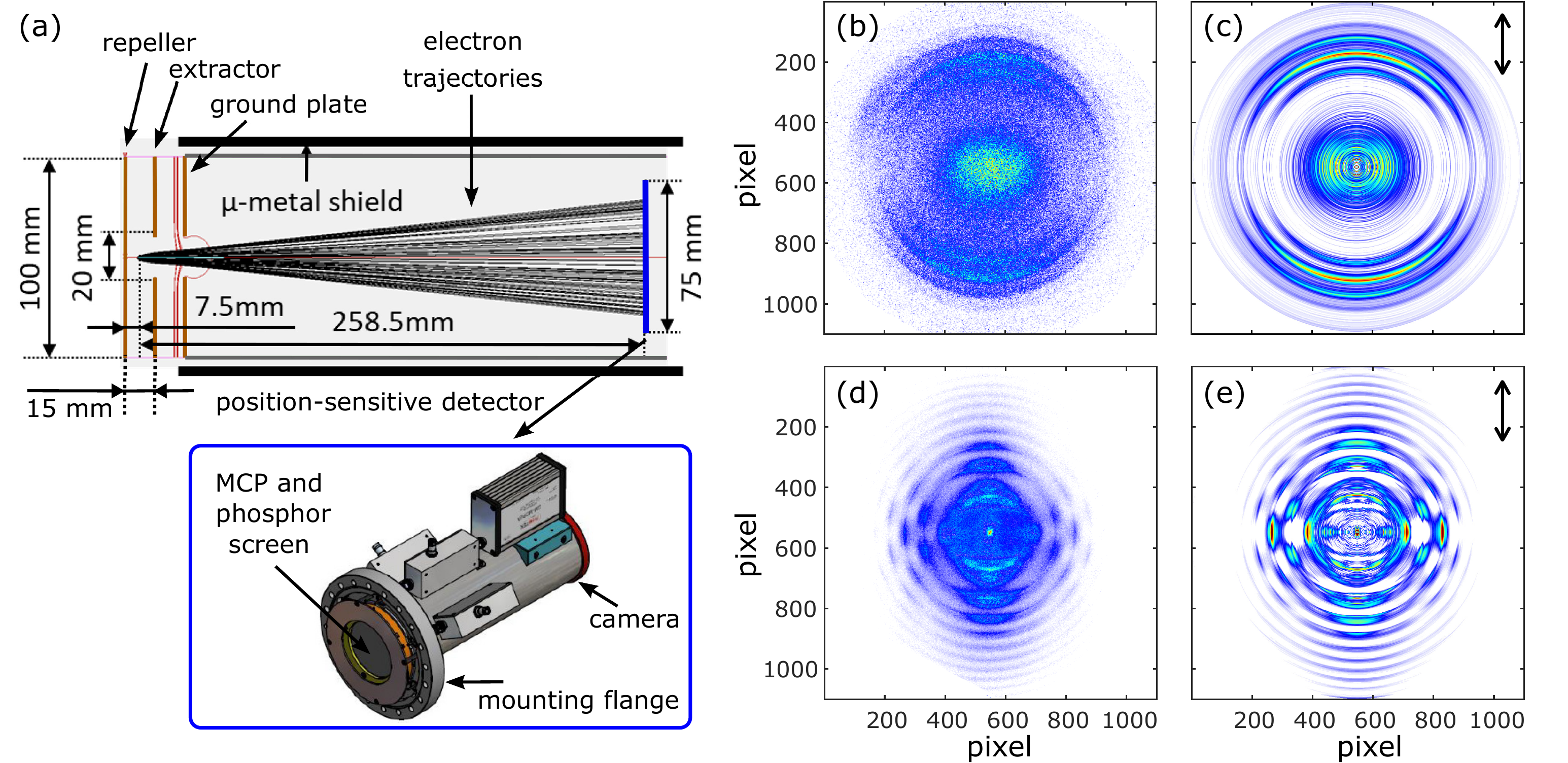}
		\caption{\label{fig_vmi} (a) Schematics of the velocity map imaging (VMI) electrostatic lens showing simulated trajectories of 20~eV electrons. Voltages on the electrodes in the simulation: $V_R=-1.8$~kV, $V_E=-1.26$~kV. The VMI lens is attached to the position sensitive detector (Photek/Velocitas), shown in the bottom panel. (b,d) Raw VMI images of electrons created by (b) single photon ionization of argon by high harmonic beam (photon energies $23-29$~eV) and (d) above-threshold ionization of xenon by the auxiliary NIR beam (800~nm, 130~fs, $2\times10^{14}$~W\,cm$^{-2}$). (c,e) Slices through 3D photoelectron momenta distributions reconstructed from (b) and (d), respectively, by the polar onion peeling algorithm \cite{roberts2009}. The laser polarization direction is indicated by arrows.} 
	\end{center}
\end{figure}

In a velocity map imaging (VMI) spectrometer the velocity vector of charged particles (ions or electrons) is mapped on the position on the detector. The VMI spectrometer at the MAC end-station is based on the design of Eppink and Parker \cite{eppink1997} and has been manufactured by Photek (Velocitas). It consists of a repeller, extractor and a ground electrode, followed by a field-free region and a position sensitive detector (Fig.~\ref{fig_vmi}(a)). For sharp velocity imaging, the voltage on the extractor is set to around 0.7-times the repeller voltage. The detector is a 75~mm diameter MCP with a P43 phosphor screen and a camera (IDS, UI-3060CP-M-GL, 166 frames per second). The MCP can be gated at a repetition rate of up to 1~kHz with a width of the temporal gate ranging from 9~ns (FWHM) up to few $\mu$s.  

The VMI spectrometer at the MAC end-station has been tested both with the high harmonic beam and the auxiliary NIR beam (Fig.~\ref{fig_vmi}(b-e)). In the first case, an argon atomic beam (section~\ref{sec_mol_beams}) was ionized by the HHG beam with photon energies in the range of $23-29$~eV (15th$- 19$th harmonic). In the second case, the MAC chamber was backfilled with xenon that was ionized by an auxiliary NIR beam (800~nm, 130~fs, $2\times10^{14}$~W\,cm$^{-2}$).
The raw photoelectron images on the camera (Fig.~\ref{fig_vmi}(b,d)) were reconstructed by the polar onion peeling algorithm \cite{roberts2009} (Fig.~\ref{fig_vmi}(c,e)). In the single-photon ionization of argon (Fig.~\ref{fig_vmi}(b,c)), three rings, corresponding to electrons from Ar $3p$ orbital released by three distinct harmonics (15th, 17th and 19th), are well resolved in the reconstructed image. The noise in the center of the reconstructed image comes from the chamber background and from numerical errors accumulated at small radii \cite{roberts2009}. In the above-threshold ionization of xenon (Fig.~\ref{fig_vmi}(d,e)), rings spaced by a photon energy of 1.55~eV are visible. Moreover, a pattern arising from interference of different quantum trajectories \cite{korneev2012} is observed. These results confirm the good performance of the VMI spectrometer.

\subsection{Magnetic bottle electron spectrometer}

\begin{figure}
	\begin{center}
		\includegraphics[width=8.5cm]{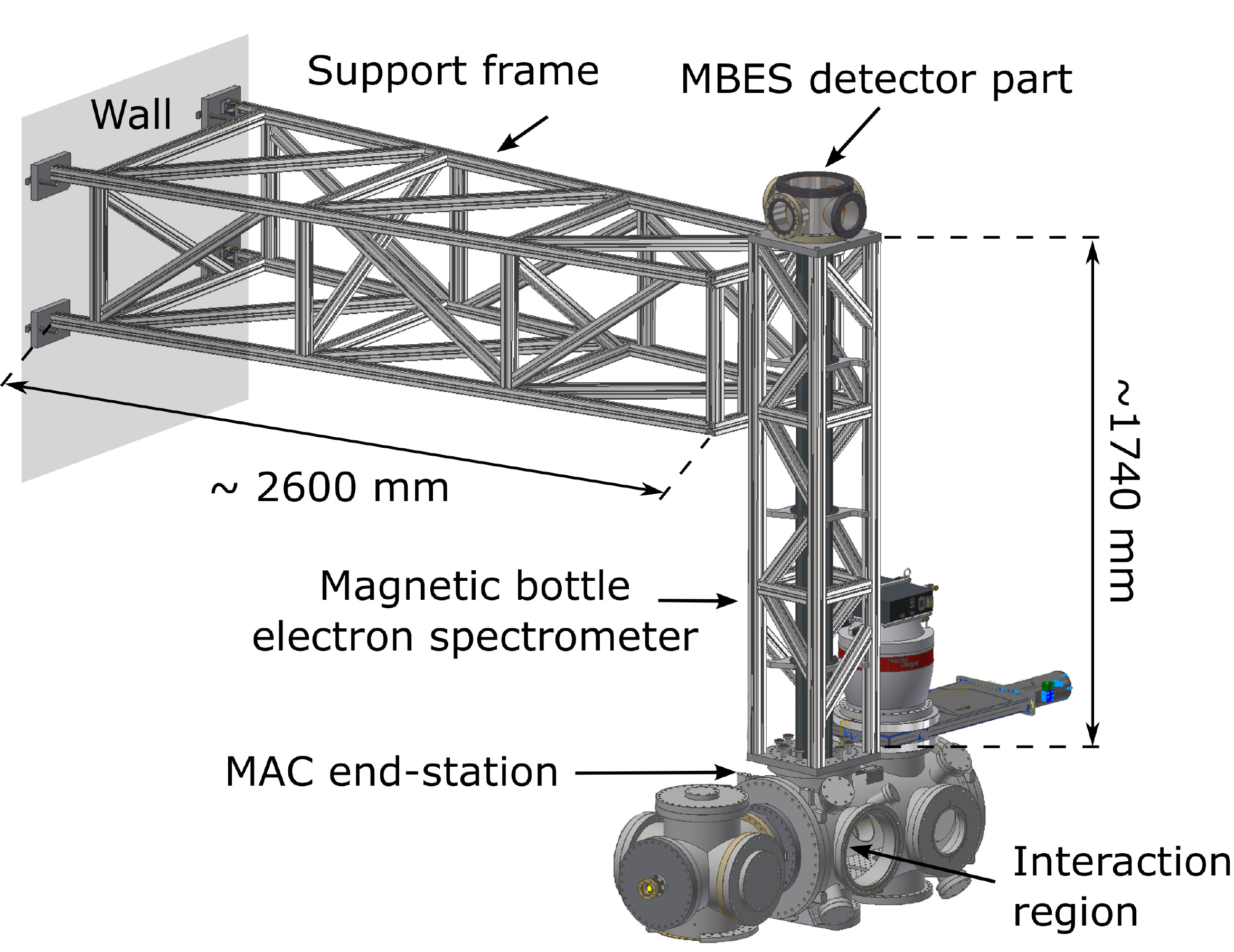}
		\caption{\label{fig_mbes} Model of the magnetic bottle electron spectrometer (MBES) installed on the top port of the MAC end-station. A frame around the MBES and a frame connecting the MBES to the wall are also shown.}
	\end{center}
\end{figure}

A magnetic bottle electron spectrometer (MBES) is under development for the MAC end-station. The design of the MBES (Fig.~\ref{fig_mbes}) is similar to that described by Eland \textit{et al.} \cite{eland2003} with an about 2~m long time-of-flight tube. The MBES will provide a very high collection and detection efficiency of essentially all electrons emitted into a solid angle of 4$\pi$ in the interaction region over a large range of electron energies \cite{roos2016}. With the implementation of a funnel type MCP detector (Hamamatsu, model F9892-31) with an open area ratio of >90\%, the electron detection efficiency has the potential to reach 90\% for electrons with energies ranging from near-zero to about 200~eV. The MBES should provide an energy resolution of about $\Delta E = E/50$. It will be possible to mount the MBES in either vertical or horizontal orientation onto the MAC chamber to accommodate for different configurations.

It will also be possible to operate the MBES as an electron-ion coincidence spectrometer when combined with a Wiley-McLaren type mass spectrometer \cite{eland2006}. The expected electron energy resolution in this configuration is around $\Delta E = E/20$ \cite{roos2018}. The ion collection efficiency may reach 50\% \cite{roos2018}, giving an overall collection and detection efficiency of about 45\%. Thus, the MBES will provide an efficient detection tool for experiments that benefit from high collection efficiency, very good spectral resolution and simultaneous detection of ions and electrons.

\subsection{Photon imaging detectors}

\begin{figure}[t]
	\begin{center}
		\includegraphics[width=15cm]{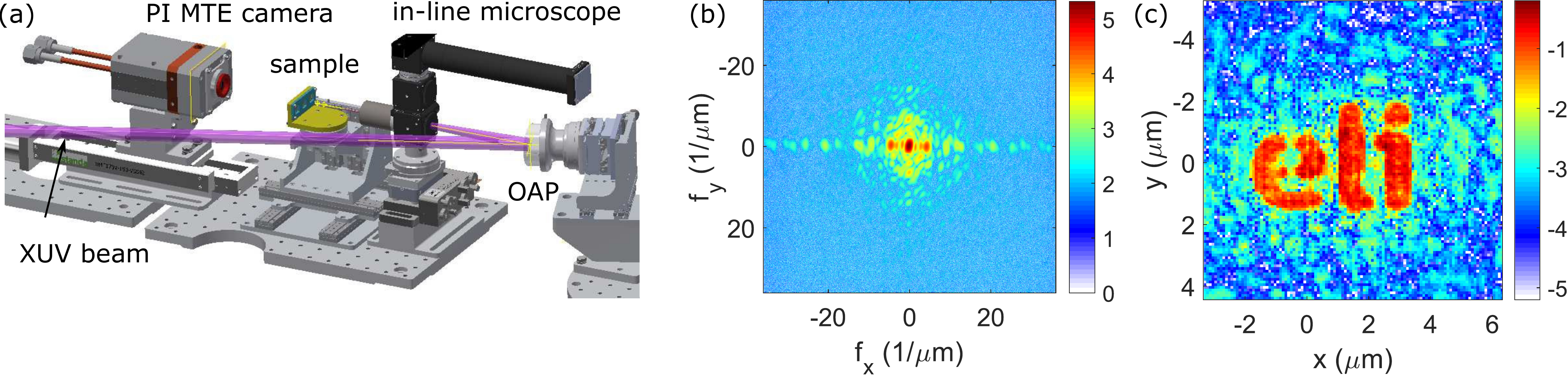}
		\caption{\label{fig_cdi} (a) Setup for coherent diffractive imaging of solid samples in the back-focusing geometry. The sample surface is observed by the in-line microscope. (b) Raw diffraction image measured from a solid sample resembling the ELI logo irradiated by the 21st harmonic selected by the monochromator. (c) Image reconstructed from the diffraction pattern in (b).}
	\end{center}
\end{figure}

Besides ion and electron spectroscopy, XUV photons scattered from the sample can be detected at the MAC end-station. For CDI experiments, an in-vacuum back-illuminated CCD camera is available (PI-MTE, Princeton Instruments, $2048\times2048$ imaging array, 13.5~$\mu$m pixel size, 100\% fill factor, photon energy range 1~eV $- 10$~keV). 
Two configurations are possible: (i) a forward focusing geometry (Fig.~\ref{fig_focusing}(a)) for single-harmonic CDI with a monochromatized beam and (ii) a back-focusing geometry with a coated OAP (Fig.~\ref{fig_cdi}(a)), allowing CDI with multiple harmonics depending on the reflectivity of the specific coating. 
In both geometries, the sample is mounted on a piezo-driven stage (providing xyz movement and rotation around the vertical axis). Multiple samples can be mounted on the sample holder. An in-line microscope can be used to monitor the sample surface and optimize the focal spot. This microscope has an objective lens with a hole, allowing propagation of the XUV beam through the hole, while observing the sample surface.

The feasibility of single-harmonic CDI in the forward focusing geometry (i) has been verified experimentally \cite{nejdl2021}. A sample resembling the ELI logo with a size of $5\times4\;\mu\rm{m}^2$ was illuminated by the 21st harmonic (photon energy 32.9~eV, wavelength 37.7~nm) selected by the monochromator. A coherent diffraction pattern (Fig.~\ref{fig_cdi}(b)) was captured by the PI-MTE camera. The image was reconstructed using a phase retrieval algorithm based on Fienup’s hybrid input-output \cite{fienup1978,bauschke02} combined with Luke’s relaxed-averaged-alternating-reflections algorithm \cite{luke2004}. Optimized retrieval was achieved after 35 iterations, resulting in good agreement between the reconstructed image  (Fig.~\ref{fig_cdi}(c)) and the SEM image of the sample (shown in \cite{nejdl2021}).

Other experimental geometries for CDI experiments can be investigated together with the user community. CDI experiments with high-harmonic source on fixed samples \cite{malm2020,vodungbo2012} or substrate-free nanoscale objects \cite{rupp2017} are foreseen.

\section{Conclusions}

The MAC end-station at the ELI Beamlines facility is a modular station for AMO sciences and CDI, exploiting an XUV beam produced by high harmonic generation. To enable a pump-probe capability, a synchronized auxiliary beam, which can cover a broad range of the electromagnetic spectrum, is available. The station is equipped with state-of-the art instruments for delivery of low-density samples and jets: atomic, molecular and cluster beams, liquid jets, injected organic or inorganic nanoparticles, and with advanced detectors for ions and electrons as well as scattered photons. The commissioning experiments have demonstrated the functionality of electron and ion spectrometers, of the molecular and cluster beam sources, the aerosol injector and the ability to perform pump-probe measurements. Nanoparticle injection together with ion spectroscopy has been shown. The MAC end-station is thus ready for user experiments in AMO sciences and CDI.

\section{Facility access for users}
The MAC end-station is open to users for scientific investigations in the area of AMO sciences and CDI. Access is obtained through peer-review of applications based on scientific excellence. More information on the user policy and open calls can be found at the ELI Beamlines user's web-page (\url{https://www.eli-beams.eu/users/}). Key capabilities, featured upgrades and recent research achievements are also updated at the instrument's web-page (\url{https://www.eli-beams.eu/facility/experimental-halls/e1-material-and-biomolecular-applications/mac/}). Users at ELI Beamlines can also access support labs (eg. Biolab, biochemical and chemical labs) that are equipped with advanced instruments for sample preparation and characterization, microscopes, spectrometers and controlled environment chambers (\url{https://www.eli-beams.eu/facility/laboratories-workshops/laboratories/}). 

\section{Acknowledgments}

The authors thank the research groups of Marcel Mudrich (Aarhus University, Denmark), Russell Minns (University of Southampton, United Kingdom), Thomas M\"oller (Technische Universit\"at Berlin, Germany) and Raimund Feifel (University of G\"oteborg, Sweden) as well as Thomas Gebert (Max Planck Institute, Hamburg, Germany), Bernd Sch\"utte (Max Born Institute, Berlin, Germany), Sa\v{s}a Bajt (DESY, Hamburg, Germany), Daniel DePonte and Jake Koralek (SLAC National Accelerator Laboratory, USA), Pamir Nag (J.~Heyrovský Institute of Physical Chemistry, Prague, Czech Republic), Tim Oelze (Technische Universit\"at Berlin, Germany) and Laura Dittrich for support during the commissioning phase and the contributions to the development of the MAC station through scientific discussions.

The authors are grateful to Zden\v{e}k Svoboda, Alexey Sterenzon, Kamil Kropielnicki and Martin P\v{r}e\v{c}ek for their technical help. The authors acknowledge the ELI Beamlines L1 and control system teams and Janos Hajdu for continuous support of the project. The authors thank Rachael Jack for manuscript revision.

This work was supported by the projects Advanced research using high intensity laser produced photons and particles (ADONIS) (CZ.02.1.01/0.0/0.0/16\_019/0000789), Structural dynamics of biomolecular systems (ELIBIO) (CZ.02.1.01/0.0/0.0/15\_003/ 0000447) (both from the European Regional Development Fund and the Ministry of Education, Youth and Sports) and the European Cluster of Advanced Laser Light Sources (EUCALL) via the Horizon 2020 Research and Innovation Programme under grant agreement no.~654220.

\section{Author contributions}

J.A. conceived the project. 
M.K. coordinated the development and operation of the MAC end-station. J.N. coordinated the development and operation of the HHG beamline.
E.K., O.K., Z.H., A.H.R., K.P.K., M.R., M.J., M.A., O.F., R.L., O.H., D.D.M., J.N., M.S., D.W., A.W., T.L., F.F., L.P., J.A. and M.K. contributed to the development of instruments and experimental infrastructure.
E.K., O.K., Z.H., A.H.R., K.P.K., M.J., M.A., O.F., O.H., D.D.M., R.B.F., L.B.L, F.F., L.P., J.A. and M.K. contributed to commissioning experiments.
E.K. analyzed the data and prepared figures.
The manuscript was written by E.K. with guidance from M.K. and input from O.K., Z.H., A.H.R., O.H., D.D.M., J.N., A.W. and J.A. All authors reviewed the manuscript.

\end{document}